\documentclass[
    a4paper,
    bibnotes,
    superscriptaddress,
    groupedaddress,
    showpacs,
]{revtex4}

\usepackage{xspace}
\usepackage{graphicx}
\usepackage{amssymb,amsmath}
\usepackage{amsmath}
\usepackage{fullpage, epic, eepic, psfrag}
\renewcommand{\nabla}{\triangledown}

\newcommand{\Eqref}[1]{Eq.~(\ref{#1})}
\newcommand{\Figref}[1]{Fig.~\ref{#1}}

\begin{document}

%% The title of the article is given with the usual \title command.
%%
%% Each author should be given as a separate \author command.
%%
%% For corresponding authors please use \author* and give the email
%% address as a second mandatory argument.
%%
%% The affiliation of authors is given after the authors; the
%% affiliations are numbered consecutively.
%%
%% If some authors have the same affiliation you can use the optional
%% argument of \author and \author* to give the number of that
%% The whole block is printed with the \maketitle command at the very
%% end.
%%%%%%%%%%%%%%%%%%%%%%%%%%%%%%%%%%%%%%%%%%%%%%%%%%%%%%%%%%%%%%%%%%%%%
\title{Current-induced dynamics in carbon atomic contacts}
\author{Jing-Tao L\"u}\email{jtlu@nanotech.dtu.dk}
\affiliation{DTU-Nanotech, Dept. of Micro- and Nanotechnology, Technical University of Denmark (DTU), {\O}rsteds Plads, Bldg.~345E, DK-2800 Lyngby, Denmark}
\author{Tue Gunst}
\affiliation{DTU-Nanotech, Dept. of Micro- and Nanotechnology, Technical University of Denmark (DTU), {\O}rsteds Plads, Bldg.~345E, DK-2800 Lyngby, Denmark}
\author{Per Hedeg{\aa}rd}\email{hedegard@fysik.ku.dk}
\affiliation{Niels Bohr Institute, Nano-Science Center,
University of Copenhagen, Denmark}
\author{Mads Brandbyge}\email{mads.brandbyge@nanotech.dtu.dk}
\affiliation{DTU-Nanotech, Dept. of Micro- and Nanotechnology, Technical University of Denmark (DTU), {\O}rsteds Plads, Bldg.~345E, DK-2800 Lyngby, Denmark}

%%%%%%%%%%%%%%%%%%%%%%%%%%%%%%%%%%%%%%%%%%%%%%%%%%%%%%%%%%%%%%%%%%%%%
%% The document should begin with an abstract, if appropriate. If one
%% is given and should not be, a warning is issued.
%%
%% For the three parts of the abstract, ``Background'', ``Results''
%% and ``Conclusions'', the corresponding commands should be used.
%%%%%%%%%%%%%%%%%%%%%%%%%%%%%%%%%%%%%%%%%%%%%%%%%%%%%%%%%%%%%%%%%%%%%
\begin{abstract}
%\background
The effect of electronic current on the atomic motion still poses many open
questions, and several mechanisms are at play. Recently there has been focus on the
importance of the current-induced non-conservative forces (NC) and Berry-phase
derived forces (BP) regarding the stability of molecular-scale contacts.
Systems based on molecules bridging electrically
gated graphene electrodes may offer an interesting test-bed for these effects.
%\results 
We employ a semi-classical Langevin approach in combination with DFT
calculations to study the current-induced vibrational dynamics of an atomic carbon chain
connecting electrically gated graphene electrodes. This illustrates how the device stability can be predicted solely from the modes obtained from the Langevin equation including the current induced forces. We point out that the gate offers control of the current independent of bias
voltage which can be used to explore current-induced vibrational
instabilities due the NC/BP forces.
Furthermore, using tight-binding and the Brenner potential we illustrate how
Langevin-type molecular dynamics can be performed including the Joule heating
effect for the carbon chain systems. Molecular dynamics including current-induced forces enables an energy
redistribution mechanism among the modes, mediated by anharmonic interactions, which is
found to be vital in the description of the electronic heating.
%\conclusion 
We have developed a semi-classical Langevin equation approach which
can be used to explore current-induced dynamics and instabilities.  We find
instabilities at experimentally relevant bias and gate voltages for the carbon
chain system.
\end{abstract}

%%%%%%%%%%%%%%%%%%%%%%%%%%%%%%%%%%%%%%%%%%%%%%%%%%%%%%%%%%%%%%%%%%%%%
%% Keywords can be given with the \keywords command which takes five
%% arguments. The arguments have to be sorted.
%%%%%%%%%%%%%%%%%%%%%%%%%%%%%%%%%%%%%%%%%%%%%%%%%%%%%%%%%%%%%%%%%%%%%

\maketitle
%%%%%%%%%%%%%%%%%%%%%%%%%%%%%%%%%%%%%%%%%%%%%%%%%%%%%%%%%%%%%%%%%%%%%
%% The main text starts right here. For each required and optional
%% section of the chosen document type a special command is defined.
%%
%% It is strongly recommended to use BibTeX for managing references.
%% Citations and citation lists can be given with the \cite command.
%% Please note, that not all references have been added to the
%% example document.
%%
%% For references in floats \cite is locally redefined to
%% adds the reference to the end of the list of references.
%%%%%%%%%%%%%%%%%%%%%%%%%%%%%%%%%%%%%%%%%%%%%%%%%%%%%%%%%%%%%%%%%%%%%
\section{Introduction}
The consequences of electronic current on the motion of atoms have become accentuated with the on-going
quest for molecular-scale electronics\cite{Ta.2006,MoBj.2009,SoKiJa.2009,seidemanbook11}. The atomic motion due to
electrical current is behind the long-term breakdown of interconnects
leading to failure in integrated circuits. This effect is of even greater importance for systems where
the bottle-neck for the electronic current is a few chemical bonds. The inelastic scattering by electrons on atomic vibrations leads to the well-known Joule heating, which can have impact on the electronic behavior and stability. However, recently it has been pointed out\cite{DuMcTo.2009,LuBrHe.2010,BoKuEg.2011,LuHeBr.2011} that other
current-induced forces can play a role. For instance, in the case of molecular contacts with conductance on the order of
$G_0=2e^2/h=1/12.9$k$\Omega$ ($e$ being the electron charge and $h$ Planck's constant), and under
"high" bias voltage ($\sim 1$V), the current-induced forces which does not conserve the energy of the atomic motion may lead to run-away behavior.
However, experiments in this regime are very challenging.
For example, for the typical experiments involving molecular-scale contacts between bulk electrodes
it is not possible to image the atomic structure in the presence of contact and current. Furthermore
it is highly non-trivial to add additional gate potentials in order to modify the electronic structure
 and get independent control of bias voltage and current\cite{KuDaHj.2003,SoKiJa.2009}.

%MD
On the theoretical side, it is desirable to develop computer simulations techniques such as molecular dynamics (MD), preferably without adjustable parameters, to study in detail the current-driven complex atomic processes.
To this end, we have recently developed an approach based on the semi-classical Langevin equation, which may form the basis of MD.
In this description the non-equilibrium electronic environment is described like an
effective "bath" influencing the atomic dynamics.
Especially, we identify the forces acting on the atoms due to the electronic current.
These include "extra" fluctuating forces yielding the Joule heating, a non-conservative "electron-wind" force (denoted NC), recently discussed by Todorov and co-workers\cite{DuMcTo.2009}, and a Lorentz-like force originating from the
quantum-mechanical "Berry phase" of the electronic subsystem\cite{LuBrHe.2010} (denoted BP).
The purpose of this article is two-fold. We will illustrate this semi-classical Langevin approach, and show how the current-induced effects could be investigated in molecular contacts connecting gated graphene or nanotube electrodes.

%graphene andcarbon chains
Graphene is now being explored very extensively due to its
eminent electrical and thermal transport properties\cite{GeNo.2007,NeGuPe.2009,Ge.2009}.
Besides being highly important in their own
right, carbon nanotube or graphene-based nanostructures may offer an interesting test bed for studies of current-induced effects at the atomic scale. For such systems experiments with atomic resolution, using
for instance state-of-the-art electron microscopes, can be performed in the presence of current,
allowing the dynamics to be followed down to single adatoms\cite{MeGiCr.2008}.
Electronic current has been used to induce changes in graphene-edges,
which were monitored while simultaneously passing a current through the structure\cite{JiHoMe.2009}. This was explained as
carbon edge-dimers desorbing due to Joule-heating\cite{EnFuJa.2010}. Taking this a step further one can imagine that nano-structured nanotubes or graphene can be used as an electrode interface to molecular devices based on organic chemistry\cite{GuGoHo.2008}. Especially promising prospects include the inherent 2D geometry of graphene which both enables straight-forward electrical gating, and atomic-scale imaging in the presence of current.
There has been a number of microscopy studies of single-atom carbon chains
bridging graphene\cite{MeGiCr.2008,JiLaPe.2009} or nanotubes\cite{BoBoGo.2010}.
On the theoretical side various aspects of these systems have been studied such as the formation of chains\cite{HoPoFa.2010,ErPoRo.2011},
their stability\cite{LiYuWa.2011}, and electron transport properties\cite{ChAnBe.2009,FuBrJa.2010,AkPa.2011}.
Here we will explore the current-induced forces and nano-scale Joule heating of the carbon chain system between
electrically gated graphene electrodes.

%Paper organization
The paper is organized as follows.
After a brief sketch of the semi-classical Langevin method, we will use it to study the dynamics of the carbon chain
as a function of bias and gate voltages. We point out that the gate,
which offers independent control of bias voltage and current in the system,
can be used to explore current-induced vibrational instabilities in the current-carrying chain.
Finally, we illustrate how the Langevin molecular dynamics can be performed including the Joule heating effect
for a carbon chain system, using tight-binding and the Brenner potential.

\section{Semi-classical Langevin dynamics}
\label{sec:lang}
We will here sketch the Langevin approach.
For a classical oscillator system (mass-scaled coordinate $x$) in a general non-linear force-field, $F$,
coupled linearly to a {\em bath} of harmonic oscillators it is
possible to eliminate the bath variables and describe it using the generalized Langevin equation,
\cite{ADDO.1976,Ka.2008,KaRo.2008},
\begin{equation}
	\ddot x = F(x)-\int_{t_0}^{t}\Pi^r(t,t')x(t')dt'+\xi(t)\,.
	\label{eq:gle}
\end{equation}
Here the bath is influencing the motion via two distinct force contributions, (i) a retarded time-kernel, $\Pi^r$, describing the back-action
at time $t$ after propagation in the bath due to the motion of $x$ at previous times, and (ii) a force $\xi$  of statistical nature originating from the thermal fluctuations of the bath.
In the case of thermodynamic equilibrium $\xi$ is characterized by a
temperature and is related to $\Pi^r$ by the fluctuation-dissipation theorem. Note that in general $x$, $F$, and $\xi$ are vectors and $\Pi^r$ is a matrix.
This method has been used
by Wang and co-workers\cite{Wa.2007,LuWa.2008} to describe thermal transport in the quantum limit, using phonons in the
two connecting reservoirs with different temperature as baths and by including their quantum fluctuations in $\xi$.
This reproduce the Landauer result of thermal transport in the harmonic case\cite{Wa.2007}.

It is possible to reach a semi-classical Langevin equation description of the motion of the ions coupled to the electron
gas if we assume a coupling to the electronic environment which is linear - either in the displacement from an equilibrium
or velocity (adiabatic expansion) of the ions. This Langevin/Brownian motion approach to atom scattering at metal surfaces has a
rather long history in the case of metal electrons in thermal equilibrium\cite{NO.1977,HeTu.1995}.

We have extended this to describe the dynamics of the ions in a nano-conductor between metal electrodes
in the non-equilibrium case, where an electronic current is present\cite{BrHe.1994,LuBrHe.2010}.
In order to sketch the derivation, we consider a displacement dependent tight-binding model with electron states
in the scattering region of interest $k,l$, and $H_{el}$ being the static electronic Hamiltonian
(scattering region and its coupling to left and right electrodes\cite{BrMoOr.2002}),
\begin{equation}
	H = H_{ph}(x) + H_{el} + \sum_{k,l,n}{\bf M}_{n,kl}c_k^\dagger c_l \,x_n\,.
	\label{eq:totham}
\end{equation}
Here $x$ is a
column vector made of mass-normalised displacement operator of each degrees
of freedom, e.g., $x_n=\sqrt{m_n}u_n$, $u_n$ and $m_n$ being the
displacement operator and mass,
$H_{ph}=\frac{1}{2}\dot{x}^T \dot{x}+\frac{1}{2}x^TKx$ corresponds to a
set of harmonic oscillators that couple with the electrons, $K$ being the dynamical matrix.
We imagine a localized basis-set describing the electrons in the scattering region, where $c_k^\dagger$ ($c_k$) is
the electron creation (annihilation) operator at site $k$ in this region\cite{FrPaBr.2007}.
Here we only consider the coupling to the electronic bath, but the linear coupling to an external phonon bath can be taken into account along the same lines and adds a contribution to $\Pi^r$. The derivation and
result for linearly coupled harmonic phonon bath is similar, and has been given in Ref.~\cite{Wa.2007}.
Alternatively, the dynamics of some external phonons, not coupling to the electrons directly, may be treated explicitly
in actual MD calculations, as we illustrate later (regions $DL$,$DR$ in \Figref{fig:MDchain}a).
%The electron-phonon coupling corresponds to matrix elements of the force operator ${\bf M}_{n,kl}= \langle k|\nabla_{x_n} H_{el}|l\rangle$.  We have assumed
The electron-phonon coupling corresponds to matrix elements of the force operator ${\bf M}_{n,kl}= \langle k|\triangledown_{x_n} H_{el}|l\rangle$.  We have assumed
that ${\bf M}$ is small by keeping only the term linear in $x$.

We may obtain an equation of motion for $x$ using Heisenberg's equation of motion, $\ddot x
=i[\dot x,H]$, using atomic units ($\hbar=1$) and implicit mode index ($n$),
\begin{equation}
	\ddot x = -K x - \sum_{kl}{\bf M}_{kl}c_k^\dagger c_l\equiv -K x + f_e\, .
	\label{eq:eomwe1}
\end{equation}
The latter term, $f_e$, describes the "forces" due to the interaction with the electron gas.
Importantly, these forces have a random nature\cite{Sc.1982}. We can calculate the mean value of $f_e$ by averaging it
over the non-equilibrium electronic state,
\begin{equation}
	\langle f_e \rangle = i{\rm{Tr}}[{\bf M}{\bf G}^<(t,t)]={\rm{Tr}}[(-\nabla_x H_{el})\, {\bf\rho}(t)]\,.
	\label{eq:meanfe1}
\end{equation}
Here we have introduced the electronic lesser-Greens function,
${\bf G}^<_{ij}(t,t)\equiv i\langle c_i^\dagger(t) c_j(t)\rangle$, which is equivalent to the
 density matrix, ${\bf\rho}$  (times $-i$),
and depends on $x(t)$, since the electrons are coupled to the $x$ in the Hamiltonian. This is similar to the expression
for an average force in Ehrenfest dynamics\cite{DuMcTo.2009}.

We can evaluate this perturbatively using the unperturbed electron lesser Green's function, ${\bf G}_0^<$, corresponding to
 the case of steady-state electron transport without electron-phonon interaction\cite{BrMoOr.2002},
\begin{equation}
	{\bf G}_0^{<}(\omega)=i{\bf A}_L(\omega) n_F(\omega-\mu_L) + i{\bf A}_R(\omega) n_F(\omega-\mu_R)\,,
	\label{eq:gl0}
\end{equation}
where ${\bf A}_{L/R}$ are density of state matrices for electronic states originating in the left/right electrodes, 
each with chemical potential $\mu_{L/R}$\cite{BrMoOr.2002}, which differ for finite bias voltage, $V$, as $\mu_L-\mu_R=eV$, 
and $n_F(\omega)=1/(e^{\omega/{k_BT}}+1)$ is the Fermi-Dirac distribution function. We thus treat the non-equilibrium electronic 
system as a reservoir unperturbed by the phonons. Using the non-equilibrium Greens function (NEGF) technique\cite{HaugJauhoBook2008}, 
we may write the 2nd lowest orders in $\bf{M}$ of $\langle f_e \rangle$ as,
\begin{equation}
	\langle f_e(t) \rangle \approx \int d\omega {\rm{Tr}}[(-\nabla_x H_{el})\, {\bf\rho}_0] - \int_{t_0}^{t}\Pi^r(t,t')x(t')dt'\,.
	\label{eq:meanfe}
\end{equation}
The first term yields a constant force due to the change in electronic bonding with bias and a "direct force" due to
interaction of charges with the field\cite{BeGuWiZh.2010}. Here $\rho_0=\rho_{ {\rm{eq}}}+\delta\rho$ is the 
non-equilibrium electron density matrix {\em without} electron-phonon interaction. We split it into an equilibrium
 contribution $\rho_{ {\rm eq}}$ and a nonequilibrium correction $\delta\rho$. In linear response, we get a 
 term ${\cal E}\cdot x$ from the field in $H_{el}$, ${\cal E}$ being the
external field, yielding a "direct" force involving
the equilibrium $\rho_{\rm eq}$. We also get a term involving $H_{el}({\cal E}=0)$ together with
the change in density to first order in the field $\Delta \rho\propto {\cal E}$ in the first term of \eqref{eq:meanfe}
resulting from the change of density in the chemical bonds due to the current\cite{BrStTa.2003,Br.2009}.

The second contribution is the retarded back-action of the electron gas due to the motion and is equivalent to the
retarded phonon self-energy. In the steady state, $\Pi^r$ only depends on the time difference,
and it is convenient to work in the frequency(energy) domain.
This can be expressed using the coupling-weighted electron-hole pair density of states,
$\Lambda^{\alpha\beta}$, inside or between electrodes $\alpha,\beta\in{L,R}$,
\begin{eqnarray}
	\label{eq:tltt}
	&&\Pi^r(t-t')=\int \Pi^r(\omega)e^{-i\omega t}\frac{d\omega}{2\pi},\\
&&\Pi^r(\omega)=\int  \frac{\Lambda(\omega')}{\omega'-\omega-i\delta}d\omega'\,,
\end{eqnarray}
where $\Lambda$ can be expressed in terms of the electrode DOS,
\begin{eqnarray}
\Lambda&\equiv&\sum_{\alpha\beta}\Lambda^{\alpha\beta}\,,\\
\Lambda^{\alpha\beta}_{mn}(\omega)&=&\frac{1}{\pi}\int_{-\infty}^{\infty}\frac{d\omega'}{2\pi}\;
{\rm{Tr}}[{\bf M}_m{\bf A}_\alpha(\omega'){\bf M}_n{\bf A}_\beta(\omega'-\omega)]
\left(n_F(\omega'-\mu_\alpha)-n_F(\omega'-\omega-\mu_\beta)\right).\label{eq:ehdos}
\end{eqnarray}
We have included a factor of 2 from the spin-degeneracy and have explicitly included the mode index, $m,n$ on the
coupling matrices, ${\bf M}$, and on $\Lambda$ in Eq.~\ref{eq:ehdos}.\\

The forces described by $\Pi^r_{mn}(\omega)$ in Eq.~\ref{eq:meanfe} contains a number of interesting current-induced effects. 
It is instructive to split the
kernel into parts,
\begin{equation}
\label{eq:tll}
\Pi^r_{mn}(\omega) = i\pi{\rm{Re}}(\Lambda_{mn}(\omega))
-\pi{\rm{Im}}(\Lambda_{mn}(\omega))+\pi\mathcal{H}\{{\rm{Re}}(\Lambda_{mn})\}(\omega)
+i\pi\mathcal{H}\{{\rm{Im}}(\Lambda_{mn})\}(\omega),
\end{equation}
where $\mathcal{H}\{f(x')\}(x)=\frac{1}{\pi}\mathcal{P}\int\frac{g(x')}{x'-x}dx'$ is the Hilbert transform.
The $\Lambda$ matrix has the following symmetry properties when exchanging modes($n\leftrightarrow m$) and 
electrodes( $\alpha\leftrightarrow \beta$),
\begin{equation}
\Lambda^{\alpha\beta}_{mn}(\omega)={\Lambda^{\alpha\beta}_{nm}}^*(\omega),
	\label{eq:p1}
\end{equation}
and
\begin{equation}
\Lambda^{\alpha\beta}_{mn}(\omega)=-\Lambda^{\beta\alpha}_{nm}(-\omega),
	\label{eq:p2}
\end{equation}
which are helpful when examining the terms in \Eqref{eq:tll}, which are summarized in the following:
\emph{Friction--}The first term in Eq.~\ref{eq:tll} is imaginary and symmetric in mode index $m,n$. It describes the
friction force, due to the generation of
electron-hole pairs in the electronic environment by the ionic motion. This process exists even in
equilibrium\cite{HeTu.1995}. For slowly varying ${\bf A}_{L/R}$ with energy compared to the vibrational energies(wide-band limit)
we obtain the simple time-local electronic friction force,  $-\eta_{el}\dot x$, with
 \begin{equation}
	\eta_{el}=-\frac{\pi{\rm Re}(\Lambda)}{\omega} \approx \frac{1}{2\pi} 
\sum_{\alpha,\beta}{\rm{Tr}}[{\bf M}{\bf A}_\alpha(\mu_\alpha){\bf M}{\bf A}_\beta(\mu_\beta)]
\label{eq:ehdoswb}
\end{equation}
\emph{NC (wind) force--}
The second term in Eq.~\ref{eq:tll} is real and anti-symmetric, which means that the general curl of this force is not zero. 
It is describing the NC force, discussed very recently by Dundas and co-workers\cite{DuMcTo.2009}. This force is finite, 
even in the limit of zero frequency, where the friction and Joule heating effect is not important anymore.

\emph{Renormalization--}The third term is real and symmetric and can be interpreted as a renormalization 
of the dynamical matrix. It contains an equilibrium part and a nonequilibrium correction. The equilibrium part is 
already included in the dynamical matrix if we calculate it within the Born-Oppenheimer approximation. The nonequilibrium
 part gives a bias-induced modification of the harmonic potential.

\emph{BP force--}Finally, the last term is imaginary, anti-symmetric, and
proportional to $\omega$ for small frequencies. It  can be identified as the "Berry phase" (BP) force in Ref.~\cite{LuBrHe.2010}. 
Since the direction of this force is always normal to the velocity in the abstract phase space, 
it does no work resembling a Lorentz force with effective magnetic field
\begin{equation}
	\mathcal{B}=-\pi\frac{\mathcal{H}\{ {\rm Im}(\Lambda(\omega'))\}(\omega)}{\omega}.
	\label{eq:bpb}
\end{equation}

\emph{Random forces--} The randomness of the force $f_{e}$ is characterized by its correlation
function in frequency domain,  which can again be calculated using NEGF. However,
we note that since $f_e$ is a quantum operator, $\langle f_e(t)f_e(0)\rangle$
does not result in a real number.  Instead we use the symmetrized and real
$\langle f_e(t)f_e^T(0)+ (f_e(0)f_e^T(t))^T\rangle$. This expression equals the
semi-classical result obtained from the path-integral derivation of the
Langevin equation\cite{Sc.1982,LuBrHe.2010} and reads in Fourier space,
\begin{equation}
\langle \xi_e \xi_e^T\rangle(\omega) \equiv \langle f_e f_e^T\rangle(\omega)-\langle f_e \rangle\langle f_e \rangle^T(\omega)
=-\pi\sum_{\alpha\beta} \coth\left(\frac{\omega-(\mu_\alpha-\mu_\beta)}
{2k_BT}\right)\Lambda^{\alpha\beta}(\omega)\,.
\end{equation}
This spectral power density can be used to generate an instance of the Gaussian random noise
as a function of time which is needed in MD
simulations.  Most importantly this random force contains not only the thermal excitations but
also the excess excitations leading to Joule heating\cite{BrHe.1994} via the dependence of the chemical potentials $\mu_L-\mu_R=eV$. 
Thus with this formalism it is possible to disentangle the various contributions to the forces, being either deterministic
or random in nature.

\begin{figure}
\centering
{\includegraphics[width=0.4\paperwidth]{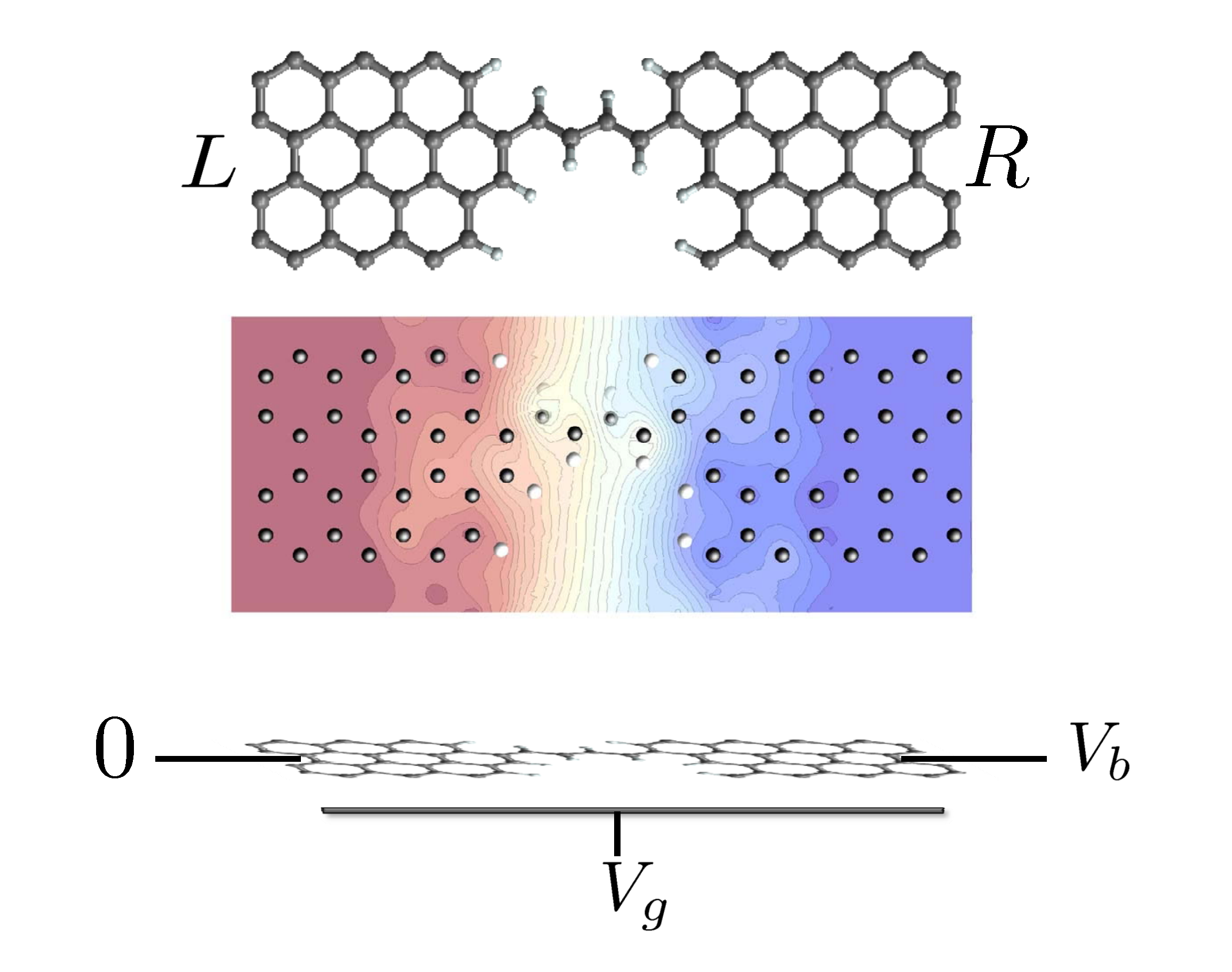}}
\caption{The system considered in the present study is a four-atom carbon chain
bridging two graphene electrodes. The dangling bonds are passivated by
Hydrogen atoms. In addition to the bias applied between the left ($L$) and
right ($R$) electrodes ($V_b$), a gate potential ($V_g$) can also be applied
perpendicular to the graphene surface. The center panel shows the calculated contour plot
of the electrostatic potential drop across the junction at $V_g=0$V, and $V_b=1$V. 
The equal drop at the left and right electrodes reflects the electron-hole symmetry for $V_g=0$V\cite{Brandbyge1999}.}
\label{fig:1}
\end{figure}

\section{Current-induced vibrational instability}
\begin{figure}
\centering
{\includegraphics[width=0.4\paperwidth]{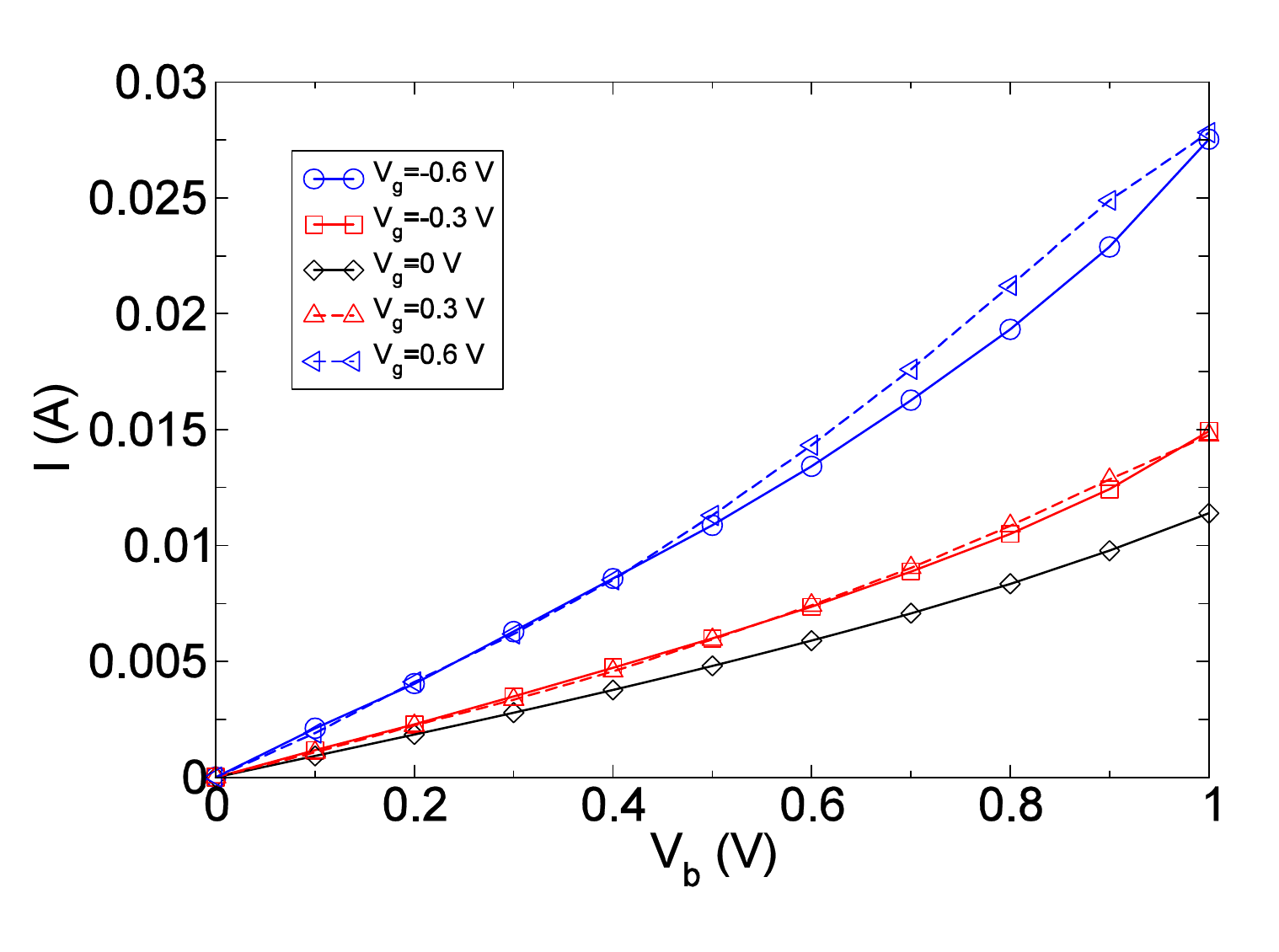}}
\caption{ Current-Voltage ($I-V_b$) curves at different $V_g$.}
\label{fig:2}
\end{figure}

We now turn to illustrations of the use of the semi-classical Langevin equation to describe current-induced effects.
In this section we employ it to study the effect
of the current-induced forces and Joule heating on the stability of the system
within the harmonic approximation. We will here ignore the coupling to electrode phonons. 
This makes an eigen-mode analysis possible which ease the interpretation of the results.
The model system we use is shown in \Figref{fig:1}, where a four-atom carbon
chain is bridged between two graphene electrodes($L$ and $R$).  We assume a
field effect transistor setup, where a gate potential, $V_g$, is applied to the
system in addition to the bias applied between the two electrodes, $V_b$. We will show that this offers a convenient way to
explore current-induced vibrational instabilities. We can already see the
effect of the gate potential in the current-voltage ($I-V_b$) characteristics
shown in \Figref{fig:2}.

\begin{figure}
\centering
{\includegraphics[width=0.4\paperwidth]{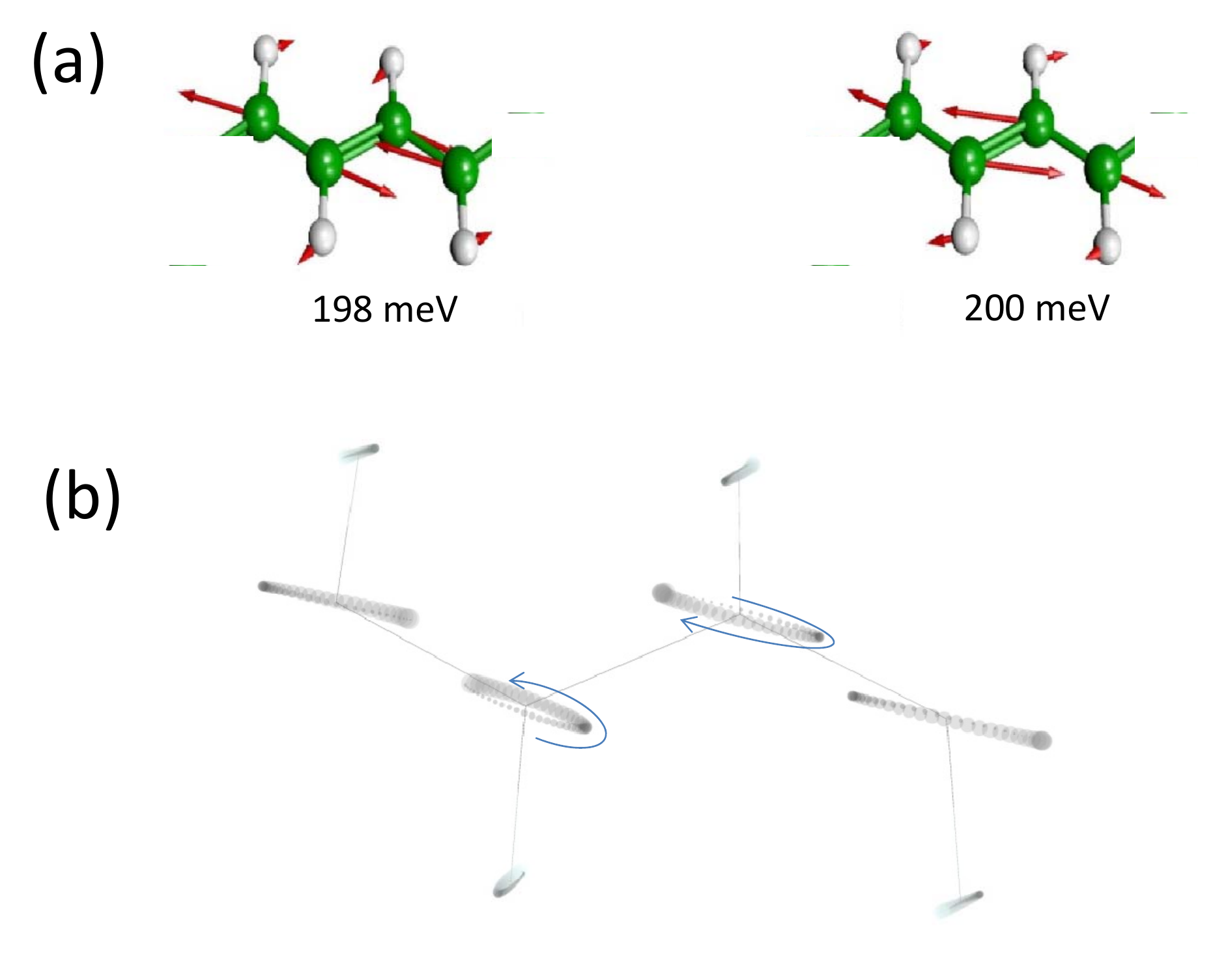}}
\caption{(a) Motion of the two phonon modes around $200$ meV. (b) Motion of the
runaway mode at $V_g=0.6$V, and $V_b=1$V.  We depict the motion using a number
of discrete time steps roughly corresponding to a full period. The position of
each atom is depicted as a circle for a sequence of time steps indicated by an increasing radius with time.
The motion is a phase-shifted linear combination of the
two modes in (a). We can see the elliptical motion of the carbon atoms from the
plot. The inclosed area indicates that work can be done by the current-induced NC force.}
\label{fig:5}
\end{figure}

The effect of the NC and BP forces is to couple different phonon modes with
nearly similar frequencies.  From now on, we will focus on the two phonon modes
around $200$ meV, shown in \Figref{fig:5}, since the alternating-bond-length-type
modes ($200$ meV) couples most strongly with the electrical current. This type of modes also gives rise to
the most intensive Raman signals in unpassivated chains between graphene-like pieces\cite{RiSaBr.2010}.

\begin{figure}
\centering
{\includegraphics[width=0.4\paperwidth]{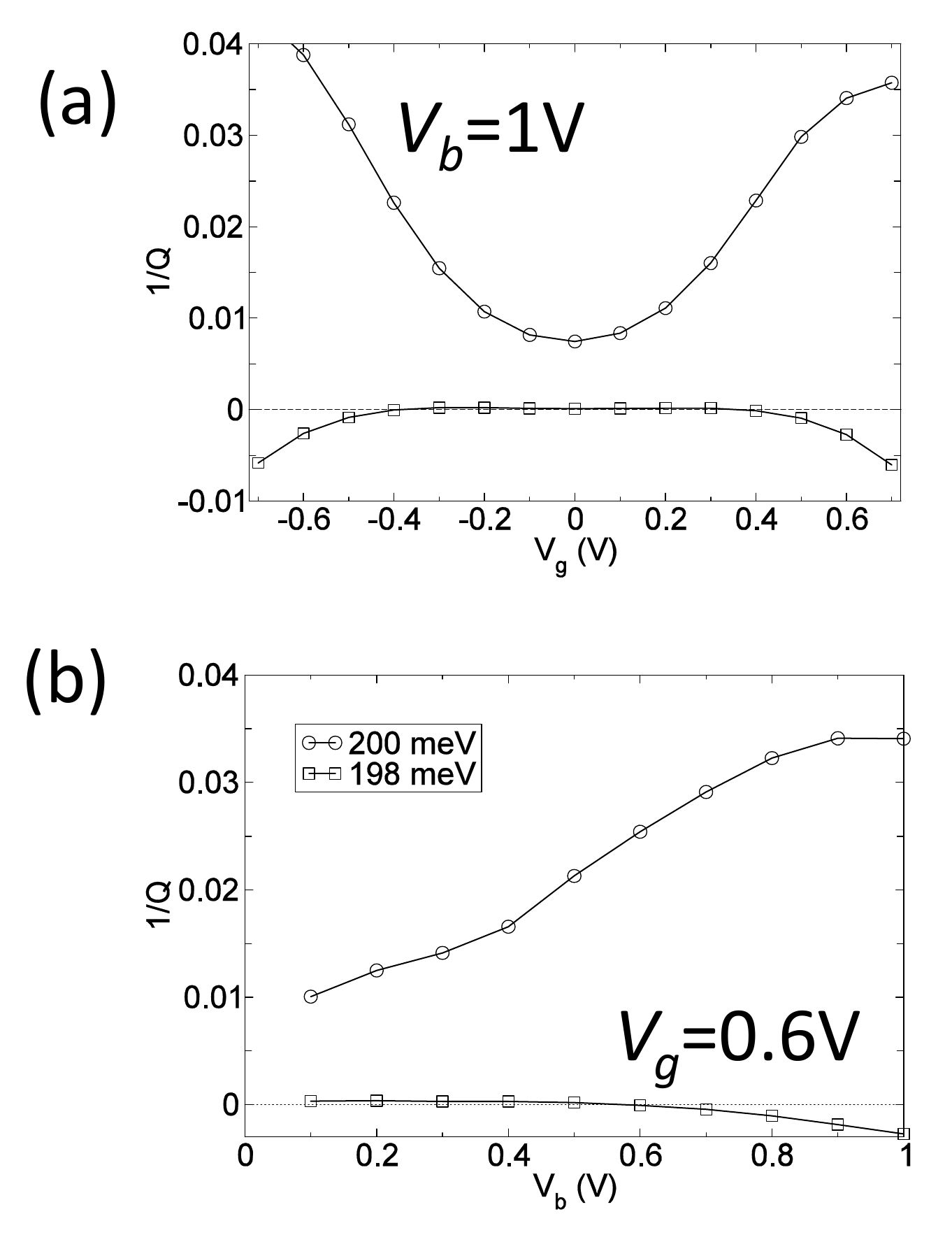}}
\caption{(a) Inverse Q-factor ($1/Q$) as a function of gate voltage, $V_g$, at $V_b=1$V for the two
modes around $200$ meV. (b) $1/Q$ as a function of bias voltage, $V_b$, at fixed gate voltage $V_g=0.6$V, for the same pair of phonon modes. }
\label{fig:3}
\end{figure}

The calculation is performed using the SIESTA density-functional theory (DFT)
method\cite{Soler.02}, which has been extended to study
elastic\cite{BrMoOr.2002} and inelastic\cite{FrPaBr.2007} transport in
molecular conductors. We use similar parameters as detailed in
Ref.~\cite{FrPaBr.2007}, and in order to keep the calculation simple and
tractable, we model the electrodes by just employing the $\Gamma$ k-point in the transverse electrode direction. 
The electron-phonon coupling matrix (${\bf M}$) is calculated at zero bias, while
we calculate the electronic structure at finite bias. We note that the voltage
dependence of the coupling matrix could play a role, but this is beyond the
present more illustrative purpose\cite{SeRoGu.2005}. Based on these approximations,
we can calculate the full $\omega$-dependent $\Lambda$ function, and the self-energies, $\Pi^r$.  To perform the
eigen-mode analysis, we further assume linear $\omega$-dependent friction, Berry
force (BP), constant non-conservative force (NC), and ignore the renormalization of the
dynamical matrix.

We model the effect of $V_b$ as a shift of the equilibrium chemical potential, $E_F$. 
In this way we can tune the electronic structure within the bias window by
changing the gate potential. In the following, we will look at the bias and gate
dependence of the inverse Q-factor ($1/Q$) and effective phonon number $N$. The
inverse Q-factor for mode $i$ (note we use index $i$ for full modes including the current-induced forces) 
is defined as
\begin{equation}
	1/Q_i \equiv -2\frac{{\rm{Im}}\{\omega_i\}}{{\rm{Re}}\{\omega_i\}},
	\label{eq:qinv}
\end{equation}
where $\omega_i$ is the eigenvalues of the full dynamical matrix, including the current-induced forces. 
These modes will then consist of linear combinations of the "unperturbed" normal modes of the system, 
$n,m$, as calculated using the standard Born-Oppenheimer approximation.
The phonon number can be calculated from the displacement correlation function,
\begin{equation}
	N_i +\frac{1}{2} \approx {\rm{Re}}\{\omega_i\} \int \langle x_i x_i \rangle(\omega) \frac{d\omega}{2\pi}.
	\label{eq:nn}
\end{equation}

\begin{figure}
\centering
{\includegraphics[width=0.4\paperwidth]{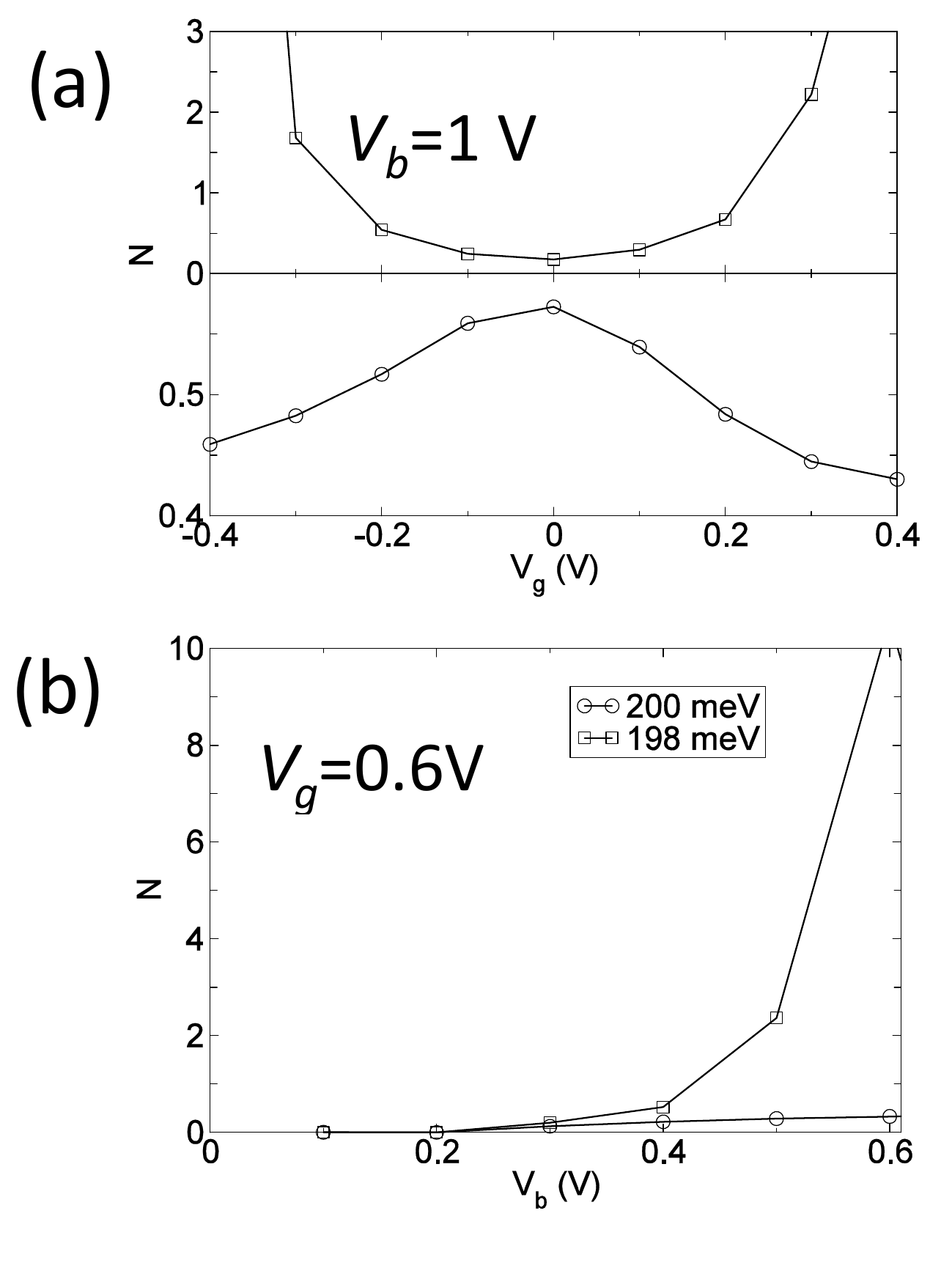}}
\caption{(a) Effective phonon number ($N$) for the two phonon modes around $200$meV as a function of gate voltage, $V_g$, at fixed bias voltage, $V_b=1$V. (b) $N$ as a function of bias voltage, $V_b$, at fixed gate voltage $V_g=0.6$V. Note that it diverges at the critical point when the damping (1/Q) in Fig.~\ref{fig:3} goes to zero. }
\label{fig:4}
\end{figure}
%1/Q and N
We show the bias and gate potential dependence of the inverse Q-factor and
phonon number in \Figref{fig:3} and \Figref{fig:4}. The coupling of these two
modes due to the bias (gate) dependent NC and BP force changes their lifetime.
The two modes always have opposite dependence. The vibrational instability
happens at the critical point where $1/Q = 0$ around $V_g = \pm 0.4$V. This
corresponds to an infinite phonon number in \Figref{fig:4}, and we therefore call it a
"runaway" mode. The motion of this mode at $V_b = 1$V, $V_g=0.6$V is plotted in
\Figref{fig:5}(b).  We can observe the elliptical motion in real-space of several atoms. This is
critical because in order for the non-conservative force to do work on the
atoms their motion has to enclose a finite area, either in real or in abstract
phase space.

Finally, we should mention that once hitting the instability threshold the
current will drive the system to some highly anharmonic regime, where present
eigenanalysis breaks down. One scenario is that the motion of the system will reach a limit cycle
 determined by the detailed anharmonic potential and the interaction with the current\cite{BoKuEg.2011}.
In this regime the detailed damping due to the coupling with phonons in the electrodes could be important,
as well as the electron-phonon coupling could change from the value around the harmonic equilibrium position.
In order to address this regime we can perform molecular dynamics simulations,
taking into account both the coupling between different modes and their
coupling with the electrode phonons, in order to study how the system actually
react due to the instability.

\section{Molecular dynamics with Joule heating}
\label{sec:MD}
%- TB model: pi, Harrison,\\
%- Whitenoise approximation + neglecting colored phonon noise.\\
%- Results: Anharmonics, Meaning and influence of \\
%- Further heating from update of coupling matrix.

\begin{figure}
\centering
{\includegraphics[width=0.6\paperwidth]{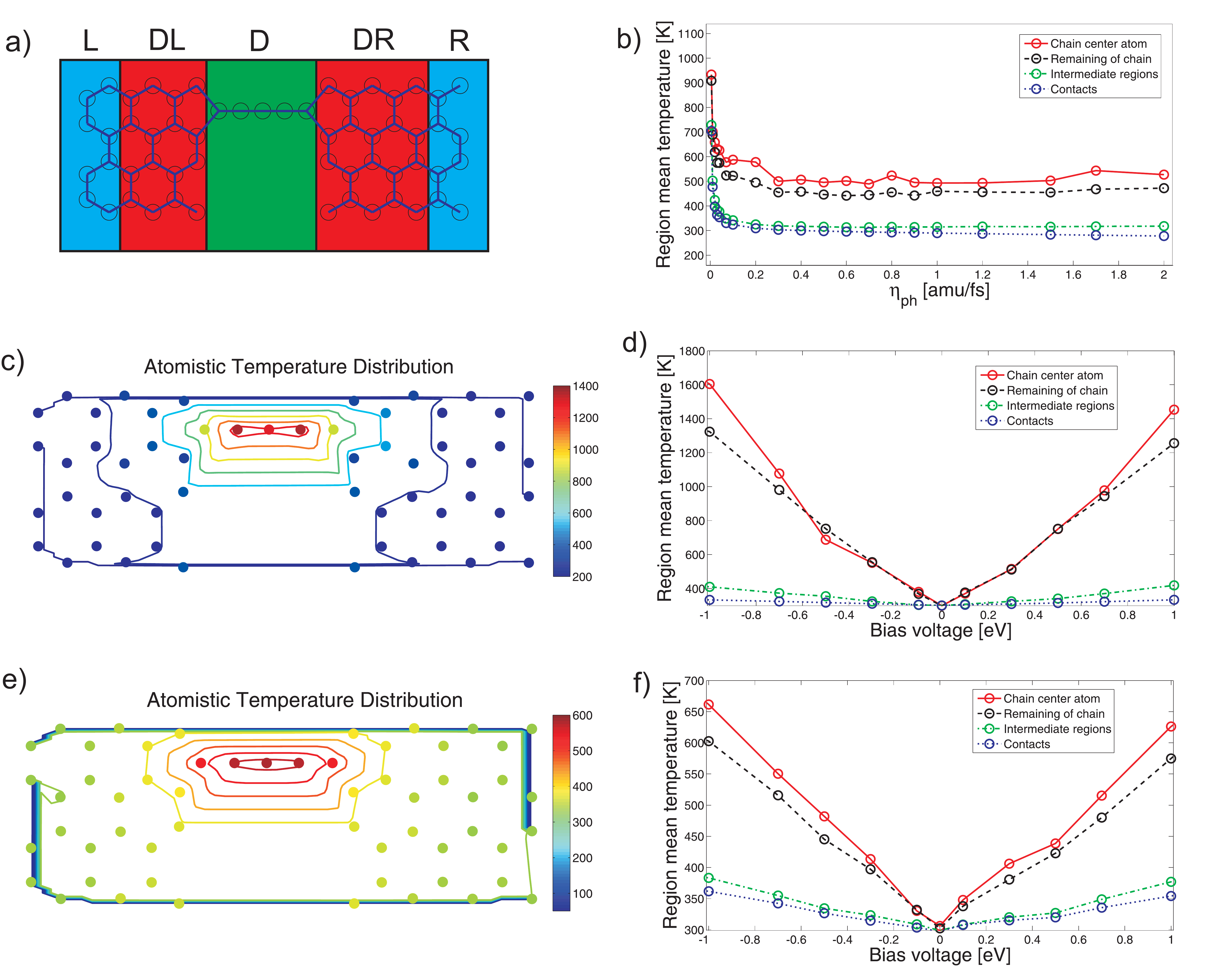}}
\caption{ (a) Definition of system regions with different types of noise contributions.
Leads ($L,R$) have a well-defined temperature specified from the phonon noise, the device ($D$) temperature
will be defined from the electronic heating, and the intermediate regions ($DL,DR$) are free and will obtain
a temperature from propagating noise. In the MD setup no atoms are held fixed but periodic boundary conditions are applied. 
The figure describes how the setup in which the local temperatures plotted in figure (c+e) should be understood. 
(b) Temperature of the regions as a function of phonon friction. (c,d)
Obtained temperatures at different atoms within the harmonic approximation. 
(c) The simulations are run at T=300K
and at $eV_b=1$eV, and (d) varying bias voltages. (e,f) Corresponding atomistic temperature distributions
including the anharmonic interactions. The lead temperature can exceed the equilibrium bath temperature due to propagating noise. 
Especially the anharmonic interactions redistribute part of the energy from the modes in the chain to the bulk modes in the lead.}
\label{fig:MDchain}
\end{figure}

Next we illustrate the use of the Langevin equation to perform molecular dynamics simulations of 
a carbon chain system in the presence of current in the simplest possible setting, but now including the coupling to electrode phonons. 
Therefore we abandon the DFT approach, and instead employ the widely used $\pi$-tight-binding model with hopping parameter 
$\beta=2.7$eV, and the Brenner potential for calculations of the inter-atomic forces\cite{brenner_empirical_1990}. 
We consider the unpassivated structure in Fig.~\ref{fig:MDchain}.  
The electron-phonon coupling is modeled by the Harrison scaling law\cite{Harrison_book}, $\beta=2.7\textrm{eV} (a_0/d)^2$, determining how
$\beta$ is modified if the nearest neighbor distance, $d$, is changed from the equilibrium value,
$a_0=1.4$\AA. The same model has recently been applied to study the effect of strain on the electronic structure of
graphene\cite{guinea_energy_2010}. In the simulation model the coupling to electrode
 phonons by a friction parameter, $\eta_{ph}$, and a corresponding white
 equilibrium phonon noise $\langle \xi_{ph} \xi_{ph}^T\rangle =2\eta_{ph}k_BT$ on the $L$,$R$ electrode regions.
 This is similar to the stochastic boundary conditions\cite{KaRo.2008}, where $L,R$-atoms act as a boundary.
 The set up for the MD is shown in (Fig.~\ref{fig:MDchain}a). We include electrode regions without 
 interaction with the current ($DL,DR$), and a device region ($D$) where the current density is highest 
and where the nonconservative forces and Joule heating is included.

Furthermore, instead of using the full non-local time-kernel for the electrons in Eq.~\ref{eq:ehdoswb}, 
we use the wide-band approximation, and neglect the off-diagonal elements of the electron noise spectral 
power density, $\langle \xi_{e} \xi_{e}^T\rangle(\omega)$.
The diagonal of the electron spectral power can be approximated by white noise in 
the high bias and wide-band limits, where variations in the electronic DOS is neglected\cite{TG_Thesis}. 
The assumption of a white noise spectrum implies neglect of
the equilibrium zero-point motion of the atoms, but most importantly here, it includes the Joule heating effects,
\begin{eqnarray}%&&
\langle (\xi_{e})_n (\xi_{e})_n^T\rangle(\omega)= 2 (\eta_{el})_{nn} k_B T +
{\textrm {Re}}\left(\textrm{Tr}\left[\mathbf{A}_{L}(\mu_{L})\mathbf{M}_{n}\mathbf{A}_{R}(\mu_{R})\mathbf{M}_{n}\right]\right) \frac{|eV_b|}{2\pi}
\end{eqnarray}
A factor of 2 should be included in the case of spin-degeneracy. Based on the velocity Verlet algorithm\cite{AllenMDnote} 
we have carried out MD simulations at
a varying voltage bias for zero gate bias ($V_g=0$V), and phonon friction, $\eta_{ph}$.
The MD results are summarized in figure \ref{fig:MDchain}(b-f). We note that for the present system set-up the 
non-conservative force is found not to play a dominant role compared to the effect of Joule heating. 
The main insight we gain from the MD example here is that the anharmonic couplings are important and effective 
in redistributing the energy supplied by the non-equilibrium electrons.

The approximate local phonon friction, $\eta_{ph}$, can in general be expressed from the slope of the corresponding
phonon self-energy at zero frequency as for electrons, see Eq.~\ref{eq:ehdoswb}. 
However, here we have simply varied its value around this in order to quantify the dependence of the local electronic heating
in the device region on this parameter (Fig. \ref{fig:MDchain}b).
The electronic heating of the chain is found not to depend much on the phonon friction when this is chosen sufficiently high.
This is an appealing result, since it indicates that the electronic heating does not depend critically on the measurement setup, 
but mainly on the nature of the actual constriction. This seems to be true as long as the heat flow away from the contacts is 
sufficient to maintain the temperature of the heat baths, and the chain acts as a bottleneck for the heat conduction.
However, we note that for heat conduction in the quantum limit it is important to go beyond the white band approximation and 
include realistic self-energies for the $L,R$ electrode phonons\cite{wang_molecular_2009}. This will be explored in future work.

Inspired by the equipartition theorem, we define a local temperature variable for the atoms (indexed by $a$) with mass, $m_a$,
\begin{equation}
T_a(t) \equiv \frac{m_a}{3 k_B} \langle\vec{v}_a^2(t)\rangle\,.
\end{equation}
Comparing the obtained temperature distributions with(Fig. \ref{fig:MDchain}c,d) and without(Fig. \ref{fig:MDchain}e,f) the
 anharmonic interactions show that anharmonic couplings between the vibrational modes has a significant influence on the 
 heat transport properties and local Joule heating of the system. The heating is less localized in
the chain due to anharmonicity. This originates from the coupling of different modes and an increased coupling to the surroundings for configurations where the atoms are displaced from their equilibrium positions. Modes localized in the chain can be heated up to very high
temperatures in the harmonic approximation. When anharmonic interactions are included the energy
is redistributed and the modes are collectively heated up.

The electron-phonon interaction is typically included through a Taylor expansion of the electronic Hamiltonian
around the equilibrium positions (Eq.~\ref{eq:totham}).
Within the time-local white noise approximation it is possible to address the effect of change of electronic 
Hamiltonian and, especially electron-phonon coupling, on the motion. This amounts to updating the friction and noise on the fly
along the path. This is possible for the simple parametrization used here. Our preliminary results based on this approximation
show that the extra noise contribution from the higher-order couplings may significantly influence the 
results and increase the electronic heating further. A method which goes beyond white noise and 
includes the change in electron-phonon coupling when the system is far from the equilibrium positions, e.g. close 
to bond-breaking, remains a challenge for the future.

\section{Conclusion}
We have developed a semi-classical Langevin equation approach, which can be used to explore
current-induced atomic dynamics and instabilities in molecular conductors. The Langevin approach can be solved in 
the harmonic approximation to obtain eigenmodes and their excitation in the presence of current, as well as used for molecular 
dynamics simulations based on the full anharmonic potential. Our simple, approximate MD simulation indicates that anharmonic 
couplings play an important role for the energy redistribution and effective heat dissipation to the electrode reservoirs. 
However, the MD is computationally very demanding beyond simplified model electronic structures and interatomic potentials and 
further developments are necessary. We have used carbon chain systems both to illustrate the Langevin approach, and in order to 
high-light how graphene might offer a unique test-bed for research into current-induced dynamic effects. Especially, it is 
straight forward to employ a gate potential to the gate electrode, and thereby obtain independent control of current and voltage 
bias in the system.
Furthermore, atomic scale resolution can be obtained in electron microscopes in the presence of current, and 
Raman spectroscopy can give insights into the excitation and effective temperature originating from the electronic 
current\cite{IoShOp.2008,WaHaCi.2008,ChKrKl.2010}. Our results for the simplified carbon chain systems indicates that 
it may be possible to tune the current-induced instabilities in the atomic dynamics with gate and bias voltages in the
 experimentally relevant range.

%%%%%%%%%%%%%%%%%%%%%%%%%%%%%%%%%%%%%%%%%%%%%%%%%%%%%%%%%%%%%%%%%%%%%
%% Tables are a bit special. Only there the \footnote is allowed for
%% Beilstein publications.
%%
%% As for figures and schemes sglcoltabular and dblcoltabular can be
%% used to get tables of the correct width. If you do not want to
%% messure the columns you can use parameter ``X'' of the tabularx
%% package for one column or more to get equal-sized columns.
%%%%%%%%%%%%%%%%%%%%%%%%%%%%%%%%%%%%%%%%%%%%%%%%%%%%%%%%%%%%%%%%%%%%%

%%%%%%%%%%%%%%%%%%%%%%%%%%%%%%%%%%%%%%%%%%%%%%%%%%%%%%%%%%%%%%%%%%%%%
%% The Supporting Information is an essential part of many articles.
%% They are given inside the ``suppinfo'' environment with the
%% \sifile command which gets the following mandatory arguments:
%% #1: File name
%% #2: File type
%% #3: Descriptive File title
%% A long description can be given using the optional argument.
%%
%% You can label each entry and reference it in the text.
%%%%%%%%%%%%%%%%%%%%%%%%%%%%%%%%%%%%%%%%%%%%%%%%%%%%%%%%%%%%%%%%%%%%%

%%%%%%%%%%%%%%%%%%%%%%%%%%%%%%%%%%%%%%%%%%%%%%%%%%%%%%%%%%%%%%%%%%%%%
%% The "Acknowledgements" section can be given in all manuscripts.
%% This should be done within the ``acknowledgements'' environment,
%% which will make the correct section title.
%%%%%%%%%%%%%%%%%%%%%%%%%%%%%%%%%%%%%%%%%%%%%%%%%%%%%%%%%%%%%%%%%%%%%
\begin{acknowledgements}
We acknowledge the Lundbeck Foundation for financial support (R49-A5454), and
the Danish Center for Scientific Computing (DCSC) for providing computer resources.
\end{acknowledgements}

\bibliography{RuitenbeekBJNano}

%%%%%%%%%%%%%%%%%%%%%%%%%%%%%%%%%%%%%%%%%%%%%%%%%%%%%%%%%%%%%%%%%%%%%
%% That's it. Ending the document finishes the article. Happy TeXing!
%%%%%%%%%%%%%%%%%%%%%%%%%%%%%%%%%%%%%%%%%%%%%%%%%%%%%%%%%%%%%%%%%%%%%
\end{document}